\documentclass[aps,prd,twocolumn,groupedaddress,floatfix]{revtex4}

\usepackage{latexsym}
\usepackage{graphicx}
\usepackage{color}
\usepackage{epsfig}

\def\lb{\left(}
\def\rb{\right)}
\def\be {\begin{equation}}
\def\ee {\end{equation}  }
\def\beq{\begin{eqnarray}}
\def\eeq{\end{eqnarray}  }
\def\bi {\begin{itemize} }

\def\ei {\end{itemize}   }
\def\RE {I\kern-6pt R    }
\def\Z  {Z\kern-13pt Z   }
\def\be {\begin{equation}}
\def\ee {\end{equation}  }
\def\beq{\begin{eqnarray}}
\def\eeq{\end{eqnarray}  }

\def\eeq{\end{eqnarray}  }

\begin{document}

\title{Threshold of Singularity Formation in the Semilinear Wave Equation}

\author{Steven L. Liebling}
    \affiliation{Department of Physics, Long Island University -- C.W. Post Campus,
                 Brookville, NY 11548}

\date{\today}

\begin{abstract}
Solutions of the semilinear wave equation are found numerically
in three spatial dimensions with no
assumed symmetry using distributed adaptive mesh refinement.
The threshold of singularity formation is studied for the two cases in which the
exponent of the nonlinear term is either $p=5$ or $p=7$.
Near the threshold of singularity formation,  numerical solutions suggest an approach
to self-similarity for the $p=7$ case and an approach to a scale evolving static solution for $p=5$.
\end{abstract}


\maketitle

\section{Introduction}

One area of interest within the context of a nonlinear wave equation is the
emergence of a singularity from smooth initial data.
Among the issues raised by the formation of a singularity, the nature of
the threshold for such formation is of particular interest. A number of past studies
have addressed this threshold numerically in the nonlinear sigma model
in both
two~\cite{2+1,Bizon2+1}
and
three~\cite{Liebling:1999nn,Bizon:2000,Liebling:2000sy,myamr,Liebling:2004nr} spatial dimensions.
Here, I study the semilinear wave equation following the work of~\cite{bizonsemi} which considers
the formation of singularities in the model restricted to spherical symmetry.

A scalar field, $\phi$, obeys the wave equation
\be
\Box \phi = \phi^p
\label{eq:box}
\ee
for $p$ odd (preserving the symmetry $\phi \rightarrow -\phi$).
In three dimensions with Cartesian coordinates, this equation becomes
\be
\ddot \phi = \phi_{,xx} + \phi_{,yy} + \phi_{,zz} + \phi^p,
\label{eq:eom}
\ee
where commas indicate partial derivatives with respect to subscripted coordinates and an
overdot indicates partial differentiation with respect to time.
Solutions are found numerically by rewriting the equation
of motion~(\ref{eq:eom}) in first differential order form and replacing
derivatives with second order accurate finite difference approximations. These finite difference equations are solved with
an iterative Crank-Nicholson scheme. In order to achieve the dynamic range
and resolution needed to resolve the features
occurring on such small scales in this model, adaptive mesh refinement
is used. In fact, the code necessary for this is
achieved with minimal modification to the code used in~\cite{myamr,Liebling:2004nr},
and the reader is referred to these earlier papers for computational details.
These changes consist of: (1)~making the association $\phi \equiv \chi$,
(2)~replacing the nonlinear term in the evolution update
of that model with the last term of Eq.~(\ref{eq:eom}), and (3)~ removing the
regularity condition on the scalar field at the origin.
As is common, an outgoing Robin boundary condition
is applied at the outer boundaries assuming a spherical outgoing front.
While not perfect, the outer boundary is generally far enough away
from the central dynamics so that the boundary condition
has no effect.

\begin{table}[h]
     \begin{tabular}{ | l | l | c | c |}
     \hline
     & Description & $\phi(x,y,z,0)$ & $\dot \phi(x,y,z,0)$ \\
     \hline
     \hline
     a & Ellipsoid            & $ G$
                          & $   \nu      \frac{\partial G}{\partial \tilde r}$\\
       &                  & ~ & $ + \Omega_z \left(y G_{,x} - x G_{,y} \right)$ \\
     \hline
     b & Two pulses   & $ G_1 + G_2 $
                          & $ v_1 \frac{\partial G_1}{\partial x} + v_2 \frac{\partial G_2}{\partial x}$ \\
     \hline
     c & Toroid                  & $ A e^{-z^2/\delta^2} \times $
                          & $ \Omega_z \left( y \phi_{,x} - x \phi_{,y} \right)$ \\
       &                         & $                    e^{ -\left(\sqrt{\epsilon_x x^2 + \epsilon_y y^2}-R\right)^2 / \delta^2 }$ & \\
     \hline
     d & Antisymmetric & $x~G$  & $0$ \\
     \hline
     e & Flat Pulse         & $A\left( \frac{1}{2} \tanh \frac{ \hat r + R}{\delta} + \frac{1}{2} \right) \times$ & $0$ \\
       &                    & $ \left( \frac{1}{2} \tanh \frac{-\hat r + R}{\delta} + \frac{1}{2} \right)$ & \\

     \hline
     f & Static             & $G + \left( 1+r^2/3 \right)^{-1/2}$  & $0$ \\
       &        (for $p=5$) &                        & \\
     \hline
     \end{tabular}
\caption{\label{table:init}List of various initial data families. 
         For families (a)-(f) both the field $\phi(x,y,z,0)$ and its
         time derivative $\dot \phi(x,y,z,0)$ are shown at the initial time $t=0$ in terms of various
         parameters. The terms $G$, $G_1$, and $G_2$ represent unique Gaussian pulses
         as defined in Eq.~(\ref{eq:id}). In family (b), the parameters
         $v_1$ and $v_2$ are the respective velocities of the two pulses, generally
         chosen to have a grazing collision. Family (f) represents perturbations of the
         static solution for the particular case of $p=5$.}
\end{table}
%
%
%

\begin{figure}
\centerline{\includegraphics[width=10cm]{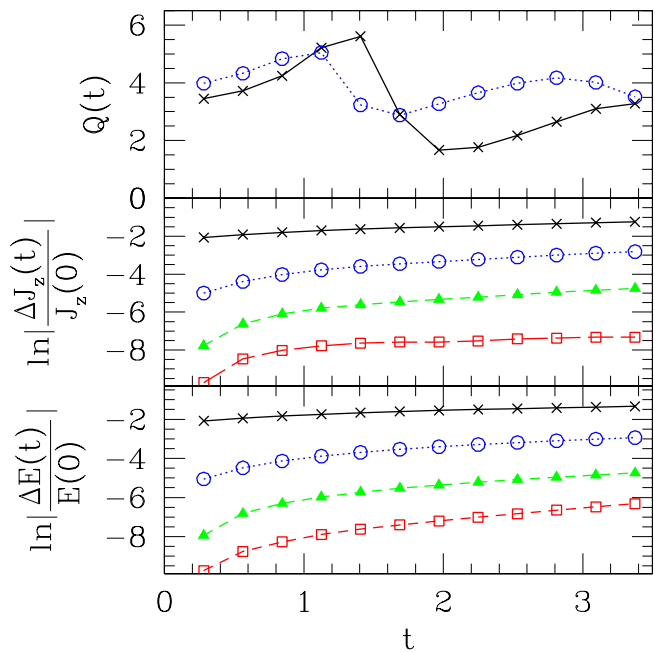}}
\caption{\label{fig:converge} Demonstration of convergence of the code for $p=7$. 
The {\bf bottom frame} displays the energy loss (relative to the initial time) $\Delta E(t) \equiv E(t) - E(0)$
for four resolutions:
$\left(32 +1\right)^3$ (solid,      black crosses),
$\left(64 +1\right)^3$ (dot,        blue circles),
$\left(128+1\right)^3$ (short dash, green triangles),
and
$\left(256+1\right)^3$ (long dash,  red squares).
As the resolution increases, energy loss decreases.
In a similar fashion the {\bf middle frame} shows
the loss of the $z$-component of the angular momentum as a function of time which
also converges to conservation.
The {\bf top frame} shows the 
convergence factor Q(t) as in Eq.~\ref{eq:cvf} which approaches the expected value of $4$
with increasing resolution.
This run is a bit below the threshold for singularity formation and, as such, represents a strong field
example away from the linear regime. The initial data is from family (c) (see Table~\ref{table:init})
with parameters $A=0.14$, $\delta=2.5$, $\epsilon_x=2.4$, $\epsilon_y=0.4$, $\Omega_z=3$, and $R=3.8$.
The dimensionless ratio of
the angular momentum to the energy squared is
$J/E^2 = 0.013$.
%
%
}
\end{figure}

Tests of the code include examining the convergence of various properties to the continuum properties as the resolution
is increased. To that end, 
the energy density associated with the scalar field is given by
\be
\rho = \frac{1}{2} \left[
                                     \lb \dot \phi \rb^2
                                   + \lb \phi_{,x} \rb^2
                                   + \lb \phi_{,y} \rb^2
                                   + \lb \phi_{,z} \rb^2
                                   \right]
                - \frac{ \phi^{p+1} }{ p+1 },
\ee
so that the energy contained in the grid can be computed by an integral.
Similarly, the $z$-component of the angular momentum is~\cite{ryder}
\be
J_z = \int d^3 x~M^{xy} = \int d^3 x ~\dot \phi \left( y \phi_{,x} - x \phi_{,y} \right).
\ee
One example of the conservation of these quantities is presented in Fig.~\ref{fig:converge}.
Also in this figure is plotted the convergence factor
\be
Q(t) = \frac{ |\tilde \chi_{4h} - \tilde \chi_{2h} |_2 }
            { |\tilde \chi_{2h} - \tilde \chi_{ h} |_2 }
\label{eq:cvf}
\ee
which compares the differences in solutions as the resolution is increased. For a second order accurate scheme 
such as this one, the convergence factor is expected to converge to the value four. As shown in Fig.~\ref{fig:converge},
the code demonstrates  second order convergence as well as conservation of the total energy and angular momentum.

\begin{figure}[ht]
\centerline{\includegraphics[width=10cm]{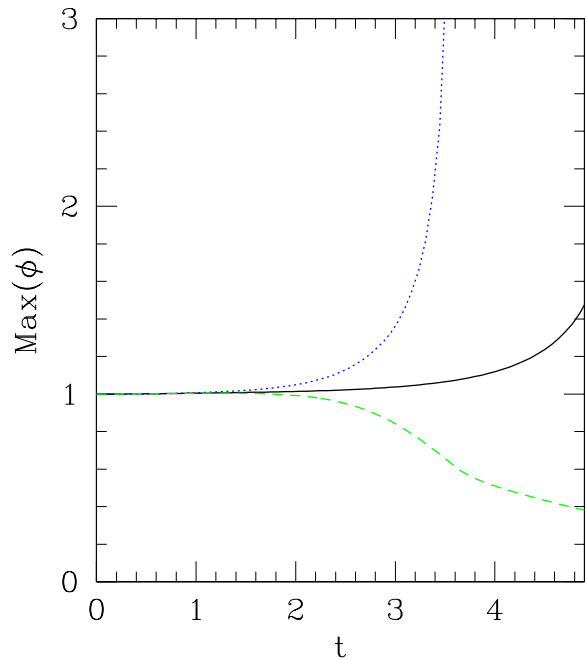}}
\caption{\label{fig:threshold_p5} Demonstration of threshold behavior of $p=5$ static solution.
The maximum value of the field is plotted for three evolutions; In solid line the results
for the static solution (Family (f) in Table~\ref{table:init}) are shown. In dashed line are the
maximums obtained with a perturbation with negative amplitude $A=-0.05$, $R=6$,
$\delta=2$, $\epsilon_x=2$, and $\epsilon_y=0.6$. In dotted line the results with the same perturbation
except with a positive amplitude $A=+0.05$ are shown. As the figure indicates, the perturbations send
the solution towards either singularity formation or dispersal.
Here, one can also see that at late times the unstable static solution is also driven to singularity
formation because of the variety of numerical perturbations inherent in the numerical scheme such
as the boundary treatment.
%
%
}
\end{figure}

A number of families of initial data have been explored, and these are described in Table~\ref{table:init}.
Initial data is created by specifying $\phi$ and $\dot \phi$ at the initial time, and is done so
here with a variety of real constants. A generalized Gaussian pulse is defined as
\be
     G(x,y,z)  =  A e^{-(\tilde r - R)^2 / \delta^2},
\label{eq:id}\\
\ee
where $\tilde r$ is a generalized radial coordinate
\be
\tilde r = \sqrt{ \epsilon_x \left(x-x_c\right)^2
                + \epsilon_y \left(y-y_c\right)^2
                +            \left(z-z_c\right)^2 }.
\ee
Such a pulse depends on parameters: amplitude $A$, shell radius $R$,
pulse width $\delta$, pulse center $\left(x_c,y_c,z_c\right)$,
 and skewing factors $\epsilon_x$ and $\epsilon_y$.
For $\epsilon_x \ne 1 \ne \epsilon_y$ such a pulse has elliptic
cross section.
Family (a) represents a single
pulse for which 
the parameter $\nu$ takes the values
$\{-1,0,+1\}$ for an approximately out-going, time-symmetric, or approximately
in-going pulse, respectively.
The angular momentum of the pulse about the $z$-axis
is proportional to the parameter $\Omega_z$ as well as to $\left(\epsilon_x - \epsilon_y\right)^2$.
The other families are similarly defined.

\section{Threshold Behavior}
Solutions of Eq.~(\ref{eq:eom}) approach one of two end states, either dispersal or singularity formation.
This situation is quite similar to the evolution of a scalar wave pulse coupled to gravity which tends toward
either dispersal or black hole formation. Choptuik's study of this problem~\cite{choptuik93} found
fascinating behavior at the threshold for black hole formation (for reviews of {\em black hole
critical phenomena} see~\cite{Choptuik:1997mq,Gundlach:1998wm}).
It was largely in the spirit of Choptuik's work that the previously mentioned studies
considered what happens at the threshold of singularity formation in the nonlinear
sigma model~\cite{Liebling:1999nn,Bizon:2000,Liebling:2000sy,myamr,Liebling:2004nr,2+1,Bizon2+1}.

Self-similar solutions are often found at the threshold, and indeed such is the case here.
As mentioned in~\cite{bizonsemi}, the scaling symmetry of Eq.~(\ref{eq:box}) allows
for self-similar solutions of the form
\be
\phi(r,t) = \left( T - t \right)^{-\alpha} U\left( \rho\right )
\label{eq:form}
\ee
where
\be
\rho = \frac{r}{T - t}
~~~~ {\rm and} ~~~~~
\alpha = \frac{2}{p-1}.
\ee
Here, $T$ is the collapse time associated with the formation of a singularity.
In~\cite{bizonsemi}, they find discrete families of solutions, $U_n\left( \rho \right)$, for $p=3$ and for $p=7$,
but find no nontrivial self-similar solutions for $p=5$. Plugging Eq.~(\ref{eq:form}) into Eq.~(\ref{eq:box}), 
one arrives at an ODE that can be solved with a standard shooting method and which duplicates the results
of~\cite{bizonsemi}.

I now consider the different cases for $p$ in turn.

\begin{figure}
\centerline{\includegraphics[width=10cm]{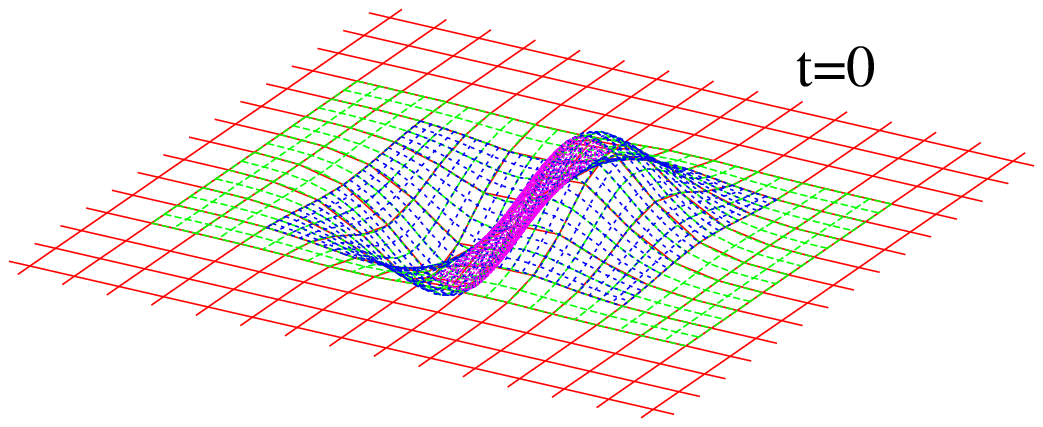}}
\vspace{-1.2in}
\centerline{\includegraphics[width=10cm]{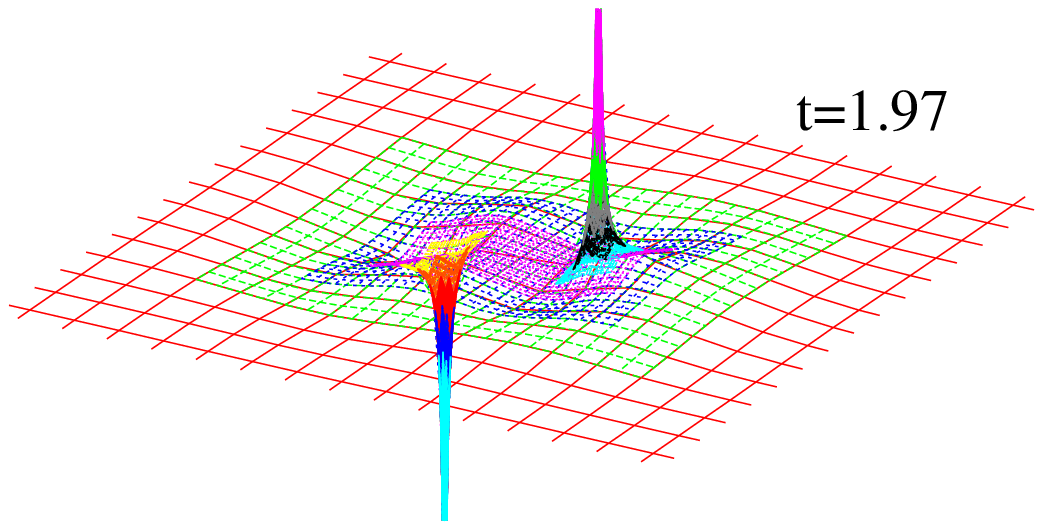}}
\vspace{-0.9in}
\caption{\label{fig:antisymm} Example of a slightly super-critical evolution of antisymmetric
initial data (Family (d)
from Table~\ref{table:init}) for $p=5$. The first frame shows the initial field configuration $\phi(x,y,0,0)$
in the $x$-$y$ plane and the second frame shows
the two, well-resolved regions in which the field is blowing up.
%
%
}
\end{figure}
%

\subsection{The case: $p=3$}
In the spherically symmetric evolutions of~\cite{bizonsemi} for $p=3$, the threshold could not be studied because
evolutions that looked to be dispersing would eventually demonstrate growth near the origin, and hence the threshold
of singularity formation could not be studied. Similar problems are encountered here even without the assumption of spherical symmetry.

\begin{figure}
\centerline{\includegraphics[width=10cm]{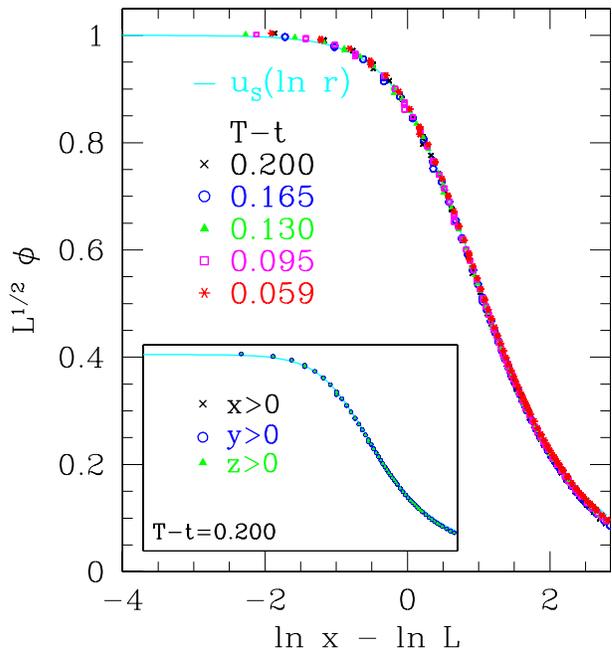}}
\caption{\label{fig:similar_p=5} Demonstration of the near critical approach to the static solution for $p=5$.
Shown are certain $x>0$, $y=0$, $z=0$ slices of an evolution at five different
times near the collapse time, $T$. Each slice $\phi(\ln x)$ is rescaled to $L^{1/2} \phi(\ln x - \ln L)$
in accordance with the rescaling symmetry of the problem. The rescaling factor $L(t)$ is chosen for each time
slice in order to achieve unity at the origin. The (unrescaled) static solution $u_S(\ln r)$ is shown in solid line.
The excellent agreement among these profiles suggests the solution represents a {\em scale evolving static solution}.
The inset shows three different slices along the three positive axes all at the same time, and indicates that
the central area of the solution is spherically symmetric.
The initial data come from family (a) of Table~\ref{table:init}
with $R=3$, $\delta=2$, $x_c=0=y_c=z_c$, $\epsilon_x=2$, $\epsilon_y=0.7$, $\nu=0$,
and $\Omega_z=0.1$, and the evolution entails $11$ levels of {\tt 2:1} refinement with a $33^3$ coarse grid.
}
\end{figure}

\subsection{The case: $p=5$}
In spherical symmetry~\cite{bizonsemi}, no nontrivial self-similar solution exists for $p=5$, and evolutions
near the threshold approach the known static solution
\be
u_S(r) = \frac{1}{\sqrt{ 1 + r^2/3}}.
\label{eq:static}
\ee
Retaining the notation of~\cite{bizonsemi}, we have a family of static solutions, $u_S^L$, generated by rescalings
of~(\ref{eq:static})
\be
u_S^L(r) = L^{-1/2} u_S \left(r/L\right).
\label{eq:staticevol}
\ee

\begin{figure}
\centerline{\includegraphics[width=10cm]{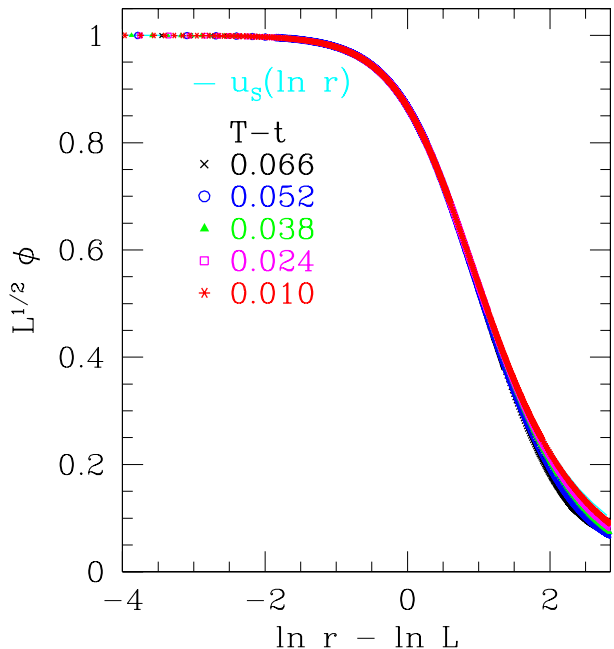}}
\caption{\label{fig:similar_ss} Near-critical solution from the explicitly spherically symmetric code for
$p=5$ with initial data family~(\ref{eq:ssform1}) and~(\ref{eq:ssform2}) and
$R=0.1$, $\delta = 0.3$, and $A^* \approx 5.539$. Similar to
Fig.~\ref{fig:similar_p=5}, the solution at times near the collapse time is shown rescaled. Its agreement
with the (unrescaled) static solution (shown in solid line) suggest a scale evolving static solution.
}
\end{figure}

In three dimensions without the assumption of spherical symmetry, the static solution remains at the threshold
of singularity formation.
In Fig.~\ref{fig:threshold_p5},
the maximum value of the scalar field within the computed domain is plotted versus time. For the static solution,
this value remains essentially constant until late times when inherent numerical perturbations drive it to
form a singularity. Also plotted are the results of initial data consisting of the static solution along with a small Gaussian pulse added
explicitly to perturb the solution. For positive amplitude of the Gaussian pulse, the solution is driven to
singularity formation whereas for negative amplitude the solution disperses. These results suggest that the
static solution remains on threshold even without the assumption of spherical symmetry.

\begin{figure}[ht]
\centerline{\includegraphics[width=10cm]{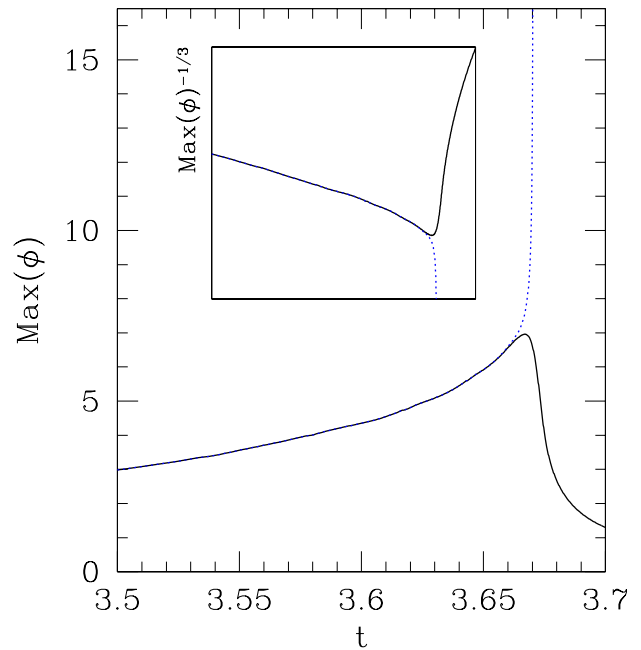}}
\caption{\label{fig:blowup} Results from two near critical evolutions for $p=7$ with initial data family (e).
The maximum of the field configuration $\phi$ is plotted versus time for a slightly subcritical evolution (solid) and
for a slightly super-critical evolution (dotted). The collapse time $T$ is estimated by determining the lifetime
of the supercritical evolution closest to criticality, in this case $T\approx 3.67$. The inset shows the same information
with the expected scaling of $\phi$.
%
%
}
\end{figure}

However, for other families of initial data near the threshold, the evolution does not appear static.
Instead, the collapsing region appears roughly self-similar in its collapse  about some central point.  Consider for example
a family of initial data which is antisymmetric across the $y$-$z$ plane, Family (d) from Table~\ref{table:init}.
One such example is shown in Fig.~\ref{fig:antisymm}. The first frame in the figure shows the initial configuration
for $\phi$ and
the second frame shows the solution near the collapse time. Two regions of  collapse form with
both regions spherically symmetric about their respective centers. 

That the collapse appears self similar when no such solution is admitted also occurs in the case of blowup with a 
Yang-Mills field~\cite{Bizon:2001,Bizon:2002}. There, the dynamics were identified with a {\em scale evolving
static solution}, and such an identification appears to be the case
here.

Evidence that the near critical solution represents the static solution with a scale factor dependent on time, $L(t)$,
is
presented for a particular case in Fig.~\ref{fig:similar_p=5}.
Shown in the figure is the evolution at different times near the collapse time rescaled according
to Eq.~(\ref{eq:staticevol}) by choosing $L(t)$ such that the rescaled quantity $\sqrt{L} \phi(r=0)$ is
unity.  The (unrescaled) static solution, $u_S$, is also shown and the excellent agreement among the profiles
is strong evidence that indeed the evolution is proceeding along the scale evolving static solution.

The inset of Fig.~\ref{fig:similar_p=5} shows three different spatial slices of
the data for a particular time. They agree quite well, providing good evidence that
the solution is spherically symmetric.

One can observe similar behavior in spherical symmetry
by modifying the code from~\cite{Liebling:1999nn}.
For many families of initial data, near threshold solutions approach the static solution in the conventional way.
However, in Fig.~\ref{fig:similar_ss} a near critical solution is shown obtained by tuning the initial
data family
\be
     \phi(r,0) = \frac{A}{2}  \left[ \tanh \left( \frac{R-r}{\delta^2} \right) + 1 \right]
	  \label{eq:ssform1}
\ee
with the initial time derivative set consistent with Eq.~(\ref{eq:form})
\be
\dot \phi(r,0) = \frac{\phi(r,0)}{2} + r \frac{d \phi(r,0)}{dr}.
	  \label{eq:ssform2}
\ee
For this family, near threshold solutions also appear to approach a scale evolving static solution.

\begin{figure}
\centerline{\includegraphics[width=10cm]{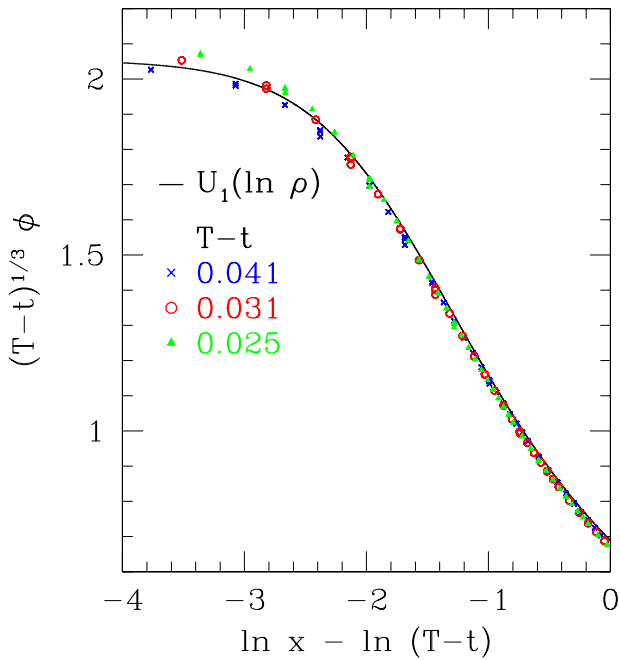}}
\caption{\label{fig:similar} Demonstration of self-similarity of a near-critical solution
for $p=7$. This is similar to Fig.~\ref{fig:similar_p=5} and indeed has the same initial data
family (with different critical amplitude $A^*$).
That the solutions at different times $t$ coincide indicates self-similarity. Also
shown is the explicitly self-similar solution $U_1(\rho)$ found by solving an ODE.
The collapse time $T$ is determined by the average of the values required for each time slice to
achieve agreement with $U_1$ for $\rho=0$.
}
\end{figure}

\subsection{The case: $p=7$}

In~\cite{bizonsemi} for the $p=7$ case, they find in the critical limit an approach to the $U_1(\rho)$
self-similar solution.  Here, without the assumption of spherical symmetry, similar threshold behavior is
observed (see~Fig.~\ref{fig:blowup}). Such a near critical solution is shown in Fig.~\ref{fig:similar}.
The solution appears both self-similar and spherically symmetric.  The inset of Fig.~\ref{fig:similar}
compares the obtained solution to the ODE solution, $U_1(\rho)$, and they appear quite similar suggesting
that it remains the critical solution even without the assumption of spherical symmetry.

\section{Conclusions}
The semilinear wave equation represents perhaps the simplest
nonlinear generalization of the linear wave equation, and yet it
displays interesting threshold behavior. In particular, this work
has extended the results of~\cite{bizonsemi} obtained in spherical
symmetry to the full 3D case.

For the $p=3$ case, the threshold could not 
be studied because of its late time growth which was also observed in
spherical symmetry.

For the $p=5$ case, a critical solution is observed which appears
roughly self similar despite the fact that no such solution is admitted.
Instead, the solutions suggest an approach to the static solution with
an evolving scale. This same behavior is also observed in a code
which explicitly assumes spherical symmetry.

For the $p=7$ case, a critical solution is found which resembles that
found in the spherically symmetric case.

\section*{Acknowledgments}
This research was supported in part by NSF cooperative agreement ACI-9619020 through computing
resources provided by the National Partnership for Advanced Computational Infrastructure at
the University of Michigan Center for Advanced Computing and
by the National Computational Science Alliance under PHY030008N.
This research was also supported in part by NSF grants PHY0325224 and PHY0139980 and by
Long Island University.

\bibliography{paper}

\begin{thebibliography}{14}
\expandafter\ifx\csname natexlab\endcsname\relax\def\natexlab#1{#1}\fi
\expandafter\ifx\csname bibnamefont\endcsname\relax
  \def\bibnamefont#1{#1}\fi
\expandafter\ifx\csname bibfnamefont\endcsname\relax
  \def\bibfnamefont#1{#1}\fi
\expandafter\ifx\csname citenamefont\endcsname\relax
  \def\citenamefont#1{#1}\fi
\expandafter\ifx\csname url\endcsname\relax
  \def\url#1{\texttt{#1}}\fi
\expandafter\ifx\csname urlprefix\endcsname\relax\def\urlprefix{URL }\fi
\providecommand{\bibinfo}[2]{#2}
\providecommand{\eprint}[2][]{\url{#2}}

\bibitem[{\citenamefont{Isenberg and Liebling}(2002)}]{2+1}
\bibinfo{author}{\bibfnamefont{J.}~\bibnamefont{Isenberg}} \bibnamefont{and}
  \bibinfo{author}{\bibfnamefont{S.}~\bibnamefont{Liebling}},
  \bibinfo{journal}{J. Math. Phys.} \textbf{\bibinfo{volume}{43}},
  \bibinfo{pages}{678} (\bibinfo{year}{2002}).

\bibitem[{\citenamefont{Bizo\'n et~al.}(2001)\citenamefont{Bizo\'n, Chmaj, and
  Tabor}}]{Bizon2+1}
\bibinfo{author}{\bibfnamefont{P.}~\bibnamefont{Bizo\'n}},
  \bibinfo{author}{\bibfnamefont{T.}~\bibnamefont{Chmaj}}, \bibnamefont{and}
  \bibinfo{author}{\bibfnamefont{Z.}~\bibnamefont{Tabor}},
  \bibinfo{journal}{Nonlinearity} \textbf{\bibinfo{volume}{14}},
  \bibinfo{pages}{1041} (\bibinfo{year}{2001}).

\bibitem[{\citenamefont{Liebling et~al.}(2000)\citenamefont{Liebling,
  Hirschmann, and Isenberg}}]{Liebling:1999nn}
\bibinfo{author}{\bibfnamefont{S.~L.} \bibnamefont{Liebling}},
  \bibinfo{author}{\bibfnamefont{E.~W.} \bibnamefont{Hirschmann}},
  \bibnamefont{and} \bibinfo{author}{\bibfnamefont{J.}~\bibnamefont{Isenberg}},
  \bibinfo{journal}{J. Math. Phys.} \textbf{\bibinfo{volume}{41}},
  \bibinfo{pages}{5691} (\bibinfo{year}{2000}),
  \eprint[http://arXiv.org/abs]{math-ph/9911020}.

\bibitem[{\citenamefont{Bizo\'n et~al.}(2000)\citenamefont{Bizo\'n, Chmaj, and
  Tabor}}]{Bizon:2000}
\bibinfo{author}{\bibfnamefont{P.}~\bibnamefont{Bizo\'n}},
  \bibinfo{author}{\bibfnamefont{T.}~\bibnamefont{Chmaj}}, \bibnamefont{and}
  \bibinfo{author}{\bibfnamefont{Z.}~\bibnamefont{Tabor}},
  \bibinfo{journal}{Nonlinearity} \textbf{\bibinfo{volume}{13}},
  \bibinfo{pages}{1411} (\bibinfo{year}{2000}).

\bibitem[{\citenamefont{Liebling}(2000)}]{Liebling:2000sy}
\bibinfo{author}{\bibfnamefont{S.~L.} \bibnamefont{Liebling}},
  \bibinfo{journal}{Pramana} \textbf{\bibinfo{volume}{55}},
  \bibinfo{pages}{497} (\bibinfo{year}{2000}), \eprint{gr-qc/0006005}.

\bibitem[{\citenamefont{Liebling}(2002)}]{myamr}
\bibinfo{author}{\bibfnamefont{S.~L.} \bibnamefont{Liebling}},
  \bibinfo{journal}{Phys. Rev.} \textbf{\bibinfo{volume}{D66}},
  \bibinfo{pages}{041703(R)} (\bibinfo{year}{2002}), \eprint{gr-qc/0202093}.

\bibitem[{\citenamefont{Liebling}(2004)}]{Liebling:2004nr}
\bibinfo{author}{\bibfnamefont{S.~L.} \bibnamefont{Liebling}},
  \bibinfo{journal}{Class. Quant. Grav.} \textbf{\bibinfo{volume}{21}},
  \bibinfo{pages}{3995} (\bibinfo{year}{2004}), \eprint{gr-qc/0403076}.

\bibitem[{\citenamefont{Bizo\'n et~al.}(2003)\citenamefont{Bizo\'n, Chmaj, and
  Tabor}}]{bizonsemi}
\bibinfo{author}{\bibfnamefont{P.}~\bibnamefont{Bizo\'n}},
  \bibinfo{author}{\bibfnamefont{T.}~\bibnamefont{Chmaj}}, \bibnamefont{and}
  \bibinfo{author}{\bibfnamefont{Z.}~\bibnamefont{Tabor}}
  (\bibinfo{year}{2003}), \eprint{math-ph/0311019}.

\bibitem[{\citenamefont{Ryder}(1996)}]{ryder}
\bibinfo{author}{\bibfnamefont{L.~H.} \bibnamefont{Ryder}},
  \emph{\bibinfo{title}{Quantum Field Theory}} (\bibinfo{publisher}{Cambridge
  University Press}, \bibinfo{year}{1996}).

\bibitem[{\citenamefont{Choptuik}(1993)}]{choptuik93}
\bibinfo{author}{\bibfnamefont{M.~W.} \bibnamefont{Choptuik}},
  \bibinfo{journal}{Phys. Rev. Lett.} \textbf{\bibinfo{volume}{70}},
  \bibinfo{pages}{9} (\bibinfo{year}{1993}).

\bibitem[{\citenamefont{Choptuik}(1997)}]{Choptuik:1997mq}
\bibinfo{author}{\bibfnamefont{M.~W.} \bibnamefont{Choptuik}}
  (\bibinfo{year}{1997}), \eprint[http://arXiv.org/abs]{gr-qc/9803075}.

\bibitem[{\citenamefont{Gundlach}(1998)}]{Gundlach:1998wm}
\bibinfo{author}{\bibfnamefont{C.}~\bibnamefont{Gundlach}},
  \bibinfo{journal}{Adv. Theor. Math. Phys.} \textbf{\bibinfo{volume}{2}},
  \bibinfo{pages}{1} (\bibinfo{year}{1998}),
  \eprint[http://arXiv.org/abs]{gr-qc/9712084}.

\bibitem[{\citenamefont{Bizo\'n and Tabor}(2001)}]{Bizon:2001}
\bibinfo{author}{\bibfnamefont{P.}~\bibnamefont{Bizo\'n}} \bibnamefont{and}
  \bibinfo{author}{\bibfnamefont{Z.}~\bibnamefont{Tabor}},
  \bibinfo{journal}{Phys. Rev.} \textbf{\bibinfo{volume}{D64}},
  \bibinfo{pages}{121701(R)} (\bibinfo{year}{2001}),
  \eprint[http://arXiv.org/abs]{math-ph/0105016}.

\bibitem[{\citenamefont{Bizo\'n}(2002)}]{Bizon:2002}
\bibinfo{author}{\bibfnamefont{P.}~\bibnamefont{Bizo\'n}},
  \bibinfo{journal}{Acta Phys. Polon.} \textbf{\bibinfo{volume}{B33}},
  \bibinfo{pages}{1893} (\bibinfo{year}{2002}),
  \eprint[http://arXiv.org/abs]{math-ph/0206004}.

\end{thebibliography}

\end{document}